\documentclass[aps,unsortedaddress,12pt]{revtex4}
\begin{document}
\title{Fermions scattering in a three dimensional extreme black hole background}
\preprint{USACH}

\author{Samuel Lepe}
\email {slepe@ucv.cl}
\affiliation{Instituto de F\'{\i}sica, Facultad de Ciencias B\'{ a}sicas y Matem\'{a}ticas,
Universidad Cat\'{o}lica de
Valpara\' {\i}so, Casilla 4059, Valpara\'{\i}so, Chile.}
\author{Fernando M\'endez}
\email {fmendez@lauca.usach.cl}
\affiliation{Departamento de F\'{\i}sica, Universidad de Santiago de Chile, Casilla 307, Santiago
2, Chile.}
\author{Joel Saavedra}
\email {jsaavedr@lauca.usach.cl} \affiliation{Instituto de
F\'{\i}sica, Facultad de Ciencias B\'{ a}sicas y Matem\'{a}ticas,
Universidad Cat\'{o}lica de Valpara\' {\i}so, Casilla 4059,
Valpara\'{\i}so, Chile}  \affiliation{Academia Polit\'ecnica
Militar, Valenzuela Llanos 623, La Reina, Santiago, Chile.}
\author{Lautaro Vergara}
\email {lvergara@lauca.usach.cl}
\affiliation{Departamento de F\'{\i}sica, Universidad de Santiago de Chile, Casilla 307, Santiago
2, Chile.}

\begin{abstract}
The absorption cross section for scattering of fermions off an
extreme BTZ black hole is calculated. It is shown that, as in the
case of scalar particles, an extreme BTZ black hole exhibits a
vanishing absorption cross section, which is consistent with the
vanishing entropy of such object. Additionally, we give a general
argument to prove that the particle flux near the horizon is zero.
Finally we show that the {\it reciprocal space} introduced
previously in \cite{gm} gives rise to the same result and,
therefore, it could be considered as the space where the
scattering process takes place in an AdS spacetime. \pacs{0460,
0470, 0470D}
\end{abstract}

\maketitle

\section{Introduction}

The presence of quantum effects in gravity is known since
Hawking's discovery that a black hole (BH) can evaporate because
of such effects \cite{hawk}. The knowledge of the full theory of
quantum gravity will tell us how this process occurs and whether
it is possible that the singularities of gravity can be smoothed
out.

The full theory of quantum gravity is unknown nowadays, but there
are strong evidence suggesting that it originated, in some limit,
from string theory.  Such conjecture is supported  by results that
relate BH  properties with string theory  effects, namely, the BH
entropy and its decay rate. For example, it has been established
that the entropy of a five dimensional extreme BH  corresponds to
the degeneracy of BPS states of a string theory \cite{strom}.
Also, the decay rate of a BH agrees with the decay rate of
thermally excited strings, both being proportional to the
absorption cross section \cite{absor}.

In addition, the classical absorption cross section (or greybody
factor) for non extreme (regular) BH in four dimensions, turns out
to be proportional to the area of the horizon \cite{area}. In this
context, great amount of research has been done; regular black
holes corresponding to different metrics have been studied
semiclassically, coupling local fields to the fixed black hole
background, in dimension four \cite{D4} and higher \cite{higher}.

In three dimensions, the existence of a BH solution of the
Einstein equations \cite{btz} has opened a window into a powerful
laboratory to prove results that, in higher dimensions, are very
hard to deal with.

Results for the greybody factor and other quantities obtained from
the scattering of scalar, fermions and photon fields off such a
BH, have been found in $2+1$ dimensions \cite{birming}, \cite{ne},
\cite{DAS}. The calculation of the same quantities for extreme BTZ
black holes deserves a careful and separate analysis, because
regular and extreme BTZ black holes correspond to different
physical objects \cite{teitel} ; they are associated to different
topologies of spacetime: the non-extreme $2+1$-dimensional black
hole has the topology of the cylinder, while the extreme case has
the topology of an annulus.

The spinless relativistic particle in an extreme BTZ background
was discussed in detail in a recent paper \cite{gm}, where it was
shown that the absorption cross section for such black hole is
zero (the same result can be deduced from \cite{Keski}). From the
thermodynamic point of view, this result can be interpreted as a
signal of zero entropy, in agreement with previous results found
in the literature \cite{teitel}.

In this paper we show that the absorption cross section for
massive spin $1/2$ particles in a $2+1$-dimensional extreme black
hole is zero, as was observed in the scalar case.  We  discuss how
to construct the states at spatial infinity, where there is no
plane wave solutions. The Dirac equation is solved for a special
case where $\omega$ and $n$, the energy and azimuthal eigenvalues,
respectively, satisfy a {\it fine tuning} condition.

The paper is organized as follows. In the following section we
review the Dirac equation in a curved space time and a $2+1$-
dimensional extreme black hole.  Section III is devoted to specify
explicitly the Dirac equation in this background and to examine
their asymptotic solutions. In Section IV the flux is constructed
and the cross section is calculated;  the reciprocal space
approach is also discussed in this context. Finally, in Section V
discussion and conclusions are presented.


\section{The Dirac equation in a 2+1-dimensional black hole background}

Let us consider a three dimensional Riemann manifold with Minkowski signature $(-,+,+) $ and line element
\begin{equation}
ds^{2}=g_{\mu \nu }dx^{\mu }dx^{\nu }.
\label{1}
\end{equation}

The non-coordinate basis one-form $e^{a}=e_{\mu}^{a}dx^{\mu }$ and the affine spin connection
$\omega _{b}^{a}=\omega _{b\mu }^{a}dx^{\mu }$ are defined by \cite{egu}
\begin{eqnarray}
ds^{2} &=&e^{a}e^{b}\eta _{ab}, \label{2} \\
de^{a}+\omega _{b}^{a}e^{b} &=&0,
\label{3}
\end{eqnarray}
where $\eta _{ab}=\mbox{diag}\left( -,+,+\right) $; latin indices
denote tangent space components and greek indices stand for
components of objects defined on the manifold.

The Dirac equation for a particle with mass $m$ in the curved background (\ref{1}) is given by \cite{naka}
\begin{equation}
\gamma ^{a}E_{a}^{\mu }\left( \partial _{\mu }-\frac{1}{8}\omega _{bc\mu }
\left[ \gamma ^{b},\gamma ^{c}\right] \right) \Psi =m\Psi ,
\label{DE}
\end{equation}
where $E_{a}^{\mu }$ is the inverse triad which satisfies
$E_{a}^{\mu }e_{\mu }^{b}=\delta _{a}^{b}$ and $\delta _{a}^{b}$
stands for the identity. $\left\{ \gamma ^{a}\right\}$ are the
Dirac matrices in the tangent space defined by the Clifford
algebra
\begin{equation}
\left\{ \gamma ^{a},\gamma ^{b}\right\} =2\eta ^{ab}.
\label{cliff}
\end{equation}

\subsection{Extreme $2+1$ black hole metric}

In order to compute the absorption cross section we must solve (\ref{DE}) in the three dimensional extreme black
hole background.

In a $2+1$-dimensional spacetime the Einstein equations with
cosmological constant $\Lambda=-\ell^{-2}$ have a solution
\cite{btz}
\begin{equation}
ds^{2}=-N^{2}\left( r\right) dt^{2}+N^{-2}\left( r\right) dr^{2}
+r^{2}\left[d\phi +N^{\phi }\left( r\right) dt\right] ^{2},
\label{btz}
\end{equation}
where the  lapse $N^{2}(r)$ and shift $N^{\phi }(r)$ functions are given by
\begin{eqnarray}
N^{2}\left( r\right) &=&-M+\frac{r^{2}}{\ell^{2}}+\frac{J^{2}}{4
r^{2}}, \\
N^{\phi }\left( r\right) &=&-\frac{J}{2r^{2}}.
\end{eqnarray}
Here $M$ and $J$ are the mass and angular momentum of the black hole, respectively.

The lapse function vanishes when
\begin{equation}
r_{\pm }=r_{ex}\left[1\pm \sqrt{1-\frac{J^{2}}{M^{2}\ell^{2}}}
\right]^{\frac{1}{2}},
\end{equation}
and therefore, the solution (\ref{btz}) is defined for
$r_+<r<\infty$, $-\pi<\phi<\pi$ and $-\infty<t<\infty$.

The extreme solution  corresponds to $J^{2}=M^{2}\ell^{2}$, in
which case $r_{\pm}=r_{ex}=\ell\sqrt{M/2}$. Hence the line element
can be written as
\begin{equation}
ds_{ex}^{2}=-\left( \frac{r^{2}}{\ell^{2}}-2\frac{r_{ex}^{2}}{
\ell^{2}}\right)dt^{2}+\frac{\ell^{2}r^{2}}{\left( r^{2}-r_{ex}^
{2}\right)^{2}}dr^{2}-2\frac{r_{ex}^{2}}{\ell}dtd\phi +r^{2}d\phi ^{2}.
\label{btze}
\end{equation}

Instead of solving (\ref{DE}) with this metric, it is convenient
to define a dimensionless set of coordinates $\left\{ u,v,\rho
\right\} $ as follows
\begin{equation}
u =\frac{t}{\ell}+\phi,\,\,\,\,\,\,\,\,\,\,\,\,\,\,\,\,\,\,\,\,\,\,\,\,\,\,\,v =\frac{t}{\ell}-\phi,\,\,\,\,\,\,\,\,\,\,\,\,\,\,\,
\,\,\,\,\,\,\,\,\,\,\,\,e^{2\rho
} =
\frac{r^{2}-r_{ex}
^{2}}{\ell^{2}},
\end{equation}
where  $-\infty<\{u,v\}<\infty$, $-\infty<\rho<\infty$. In the
space $\{u \times v\}$, two points $(u_1,v_1)$ and $(u_2,v_2)$ are
identified if they satisfy $u_1=v_2$ and $v_1=u_2$, for any value
of $\rho$.

The line element (\ref{btze}) in these new coordinates reads
\begin{equation}
ds^{2}=r_{ex}^{2}dv^{2}-\ell^{2}e^{2\rho }dudv+\ell^{2}d\rho
^{2},
\end{equation}
and according to (\ref{2}) we have the triads
\begin{equation}
e^{1}=\frac{\ell^{2}~e^{2\rho }}{2~r_{ex}}~du,\,\,\,\,\,\,\,\,\,\,\,\,\,\,\,\,\,\,\,\,\,e^{2}=r_{ex}~dv-
\frac{\ell^{2}~e^{2\rho
}}{2~
r_{ex}}~du,\,\,\,\,\,\,\,\,\,\,\,\,\,\,\,\,\,\,\,\,\,\,\,\,e^{3}=\ell ~d\rho ,
\label{15}
\end{equation}
and from (\ref{3}) the connections
\begin{equation}
\omega ^{1}\,_{2}=d\rho,\,\,\,\,\,\,\,\,\,\,\,\,\,\,\,\,\,\,\,\,\,\,\,\,\,\,\,\,\,\,\omega ^{1}~_{3}=\frac{\ell~e^{
2\rho}}{2~r_{ex}}~du
+
\frac{r_{
ex}}{\ell}~dv,\,\,\,\,\,\,\,\,\,\,\,\,\,\,\,\,\,\,\omega ^{2}~_{3}=-\frac{\ell~e^{2\rho }}{2~r_{ex}}du.
\label{16}
\end{equation}
The non vanishing components of the inverse triad are
\begin{equation}
E_{1}^{u}=\frac{2~r_{ex}~e^{-2\rho
}}{\ell^{2}},\,\,\,\,\,\,\,\,\,\,\,\,\,\,\,\,\,\,\,\,\,\,\,\,E_{1}^{v}=\frac{1}{r_{ex}}=E_{2}^{v},\,\,\,\,\,
\,\,\,\,\,\,\,\,\,\,\,\,\,\,\,\,E_{3}^{\rho}=\frac{1}{\ell}.
\label{17}
\end{equation}

In the next section we write the Dirac equation in these
coordinates and will find its solution.

\section{ The Dirac equation and its solution}
\subsection{The Dirac Equation}
We choose the Dirac matrices as follows
\begin{equation}
\gamma ^{1} =-i\sigma ^{3},\,\,\,\,\,\,\,\,\,\,\,\,\,\,\,\,\,\,\,\,\,\,\,\,\,\,\,\,\,\,\gamma ^{2} =\sigma ^{1},\,\,
\,\,\,\,\,\,\,\,\,\,\,\,\,\,\,\,\,\,\,
\,\, \,\,\,\,\gamma ^{3} =\sigma ^{2},
\end{equation}
where $\sigma^i$ are the Pauli matrices. This choice satisfies
(\ref{cliff}).

The Dirac equation (\ref{DE}) can be obtained directly using
(\ref{15}) through (\ref{17}). If we write the solution as
\[\Psi(u,v,\rho)=\left(
\begin{array}{c}
{\cal{U}}(u,v,\rho) \\
{\cal{V}}(u,v,\rho)
\end{array}\right),
\]
then (\ref{DE}) becomes
\begin{eqnarray}
\left[-i\left( \frac{2 r_{ex} e^{-2\rho } }{\ell^{2}}\partial _{
u}+\frac{\partial _{v}}{r_{ex}}\right) -(\frac{1}{2\ell}+m)
\right]{\cal{U}}+\left[
\frac{\partial _{v}}{r_{ex}}-\frac{i}{\ell}\left( \partial _{
\rho }-1\right)  \right]{\cal V}&=&0,\label{DEC1}
\\
\left[\frac{\partial _{v}}{r_{ex}}+\frac{i}{\ell}\left( \partial
_{\rho }-1\right) \right]{\cal{U}}+\left[ i\left( \frac{2r_{ex}e
^{-2\rho }}{\ell^{2}}
\partial _{u}+\frac{\partial _{v}}{r_{ex}}\right) -(\frac{1}{2
\ell}+m) \right]{\cal{V}}&=&0.
\label{DEC2}
\end{eqnarray}

In order to solve this equation, let us look for solutions of the
form
\begin{equation}
\Psi \left( u,v,\rho \right) =e^{i\left( \alpha u+\beta v\right)
}
\left(\begin{array}{c}
F(\rho)\\
G(\rho)
\end{array}\right),
\label{ansatz}
\end{equation}
where $\alpha $ and $\beta $ are constants related to the angular
and temporal eigenvalues of the solution of (\ref{DE}) in the
coordinates $\left\{ t,\phi ,r\right\} $;  namely if  the solution
behaves like $e^{i(n\phi +\omega t)}$, then
\begin{equation}
\alpha=\frac{1}{2}\left( \omega \ell+n\right),\,\,\,\,\,\,\,\,\,\,\,\,\,\,\,\,\,\,\,\,\,\,\,\,\,\,\,\,\,\,\,\,\,\,\,\,\beta  =
\frac{1}{2}\left(
\omega
\ell-n\right).
\end{equation}

For the $\rho$-dependent part of the equation it is more
convenient to define $z=e^{-2\rho }$. Therefore, using the Ansatz
(\ref{ansatz}), the $z$-part of (\ref{DEC1}) and (\ref{DEC2})
becomes

\begin{eqnarray}
\left[ \frac{2 \alpha r_{ex}}{\ell^{2}}z+\frac{\beta }{r_{ex}}-
\frac{1}{2\ell }-m\right] F\left( z\right) +i\left[ \frac{\beta
}{r_{ex}}+\frac{1}{\ell}
\left( 2z\frac{d}{dz}+1\right) \right] G\left( z\right)&=&0,
\label{ED1}
\\
i\left[ \frac{\beta }{r_{ex}}-\frac{1}{\ell}\left(
2z\frac{d}{dz}+1\right) \right] F\left( z\right)-\left[ 2\frac{
\alpha r_{ex}}{\ell^{2}}z+\frac{\beta }{
r_{ex}}+\frac{1}{2\ell}+m\right] G\left(
z\right)&=&0.\label{ED2}
\end{eqnarray}
Here we have used the same notation for the functions $F$ and $G$,
independently whether they depend either on $\rho$ or $z$. The
notation will be used through out the text unless it becomes
confusing.

One can find a solution of this set of first order coupled
differential equations by rewriting them as a second order one.
For example, if we solve (\ref{ED1}) for $F(z)$ and replace this
result in (\ref{ED2}) we find that $G(z)$ satisfies a second order
differential equation. By defining the variable  $x =\alpha(
\,r_{ex}/\ell)\,z$ one finds that $G(x)$ satisfies
\begin{equation}
A(x) G^{\prime \prime }(x) +B(x)G^{\prime}(x) +C(x)G(x) =0,
\label{eqg}
\end{equation}
with
\begin{eqnarray}
A(x)  &=&(\delta-x)x^2,
\\
B(x)  &=&(2\delta-x)x ,
\\
C(x)  &=&-x^3+(\delta-\tilde{\beta})\,x^2+
\frac{1}{4}\bigg[(\tilde{\beta}+1)^2+4\delta(2\tilde{
\beta}+\delta)\bigg]x-\frac{\delta}{4}\left((2\delta +\tilde{\beta})^2-1\right),
\end{eqnarray}
where the constant $\delta$, $m_{\mbox{\scriptsize{eff}}}$ and $\tilde{\beta}$ are given by
\begin{equation}
\delta =\frac{1}{2}\left(\ell\,m_{\mbox{\scriptsize{eff}}}-
\tilde{\beta}\right),\,\,\,\,\,\,\,\,\,\,\,\,m_{\mbox{\scriptsize{eff}}}=
m + \frac{1}{2\ell}\,
,\,\,\,\,\,\,\,\,\,\,\,\,\tilde{\beta}=\frac{\ell \beta}{r_{ex}}.
\end{equation}
The same procedure yields a similar equation for $F\left(z\right)$.

Note that this equation has three singular points for $\delta\neq
0$. Two of them are regular ($0$ and $\delta$) while the other one
located at infinity is irregular (corresponding to the horizon in
radial coordinates).

In order to solve the equation we must consider two cases: $\delta
=0$ and $\delta\neq 0$. In the first case, the equation has one
regular singularity, while in the second one, it has the three
singularities. Therefore, we will be able to completely solve the
$\delta =0$ case, while the other one must be treated in an
approximate way.

The physical meaning of such cases is as follows. For s-waves, the
$\delta=0$ condition corresponds to particles with energies
satisfying the relation $\omega =2 m_{\mbox{\scriptsize{eff}}}
r_{ex} \Omega$, where $\Omega=1/\ell$ is the angular velocity of
the horizon. One  can think of  the previous condition as fixing
the energy  of the particle to the (classical) energy of a
particle with mass $m_{\mbox{\scriptsize{eff}}}$ rotating in a
circular orbit of radius $r_{ex}$  with angular velocity $\Omega$
. For waves with $n\neq 0$, the above relation holds, but now the
energy has a contribution from the angular part.

Note that this condition yields a  precise relation between the
energy and the azimuthal eigenvalue and is a {\it fine tuning}
condition that, for a generic wave, should be hard to fulfill.

In the next section we will discuss all cases in  detail.

\subsection{Solutions of Dirac equation}

As we said previously, there exist two  distinct cases. We will
prove that, despite different solutions for both cases, they share
a common characteristic related to their behaviour near the
horizon.
\subsubsection{$\delta =0$.}

In this case, equation (\ref{eqg}) reads
\begin{equation}
x^2 G'' + x G' +\left[x^2+\tilde{\beta}x -\frac{1}{4} (\tilde{
\beta}+1)^2\right]G=0,
\label{sammy}
\end{equation}
with the following solution
\begin{equation}
G(x)=e^{-ix} x^{\frac{\tilde{\beta}+1}{2}}\left(P\,\,F\left[1
+\frac{1}{2}(1+i)\tilde{\beta},2+\tilde{\beta};2ix\right]+Q
\,\,U\left[1+\frac{1}{2}(1+i)\tilde{\beta},2+\tilde{\beta};2
ix\right]\right),
\end{equation}
where $F[c,d;x]$ and $U[c,d;x]$ are confluent hypergeometric
functions (Kummer's solution); $P$ and $Q$ are complex constants.
These functions provide a set of linearly independent solutions of
(\ref{sammy}) only if $\tilde\beta$ is non integer \cite{htf}.

The confluent hypergeometric $F[c,d;x]$ is regular at $x=0$, while
the confluent hypergeometric $U[c,d;x]$ is regular at infinity.
Thus, the regular piece of the solution at infinity (that is, near
the horizon in radial coordinate) is
\begin{eqnarray}
G(x)&=&e^{-ix} x^{\frac{\tilde{\beta}+1}{2}}Q
\,\,U\left[1+\frac{1}{2}(1+i)\tilde{\beta},2+\tilde{\beta};2
ix\right], \nonumber
\\
&\sim&e^{-ix}\sum _{n=0}^{N}\alpha_n\, x^{-n-1/2}\,\,\,\,\,\,\, \,
\,\,\,\,\,(x\rightarrow \infty),
\end{eqnarray}
where the $\alpha_n$ depend on $\tilde \beta$ and $N$ is an
arbitrary integer controlling the order of the expansion.

The regular solution for $x\rightarrow 0$ (that is, at spatial
infinity) is $F[c,d;x]$ and has a series expansion in positive
powers of $x$. However, the flux associated with this solution is
not straightforwardly defined because the BTZ spacetime is not
asymptotically flat; this is a subtle point because as was shown
in \cite{birming}. We will discuss this issue in section IV.

\subsubsection{$\delta \neq 0$.}
In this case, the equation has three singular points, being
$x\rightarrow\infty$ of irregular type. We can write (\ref{eqg})
in a more transparent way by defining the variable $y=x/\delta$
and
\begin{equation}
G(y)=\frac{1-y}{\sqrt{y}} D(y).
\end{equation}

The equation for $D(y)$ turns out to be
\begin{eqnarray}
y^2 (1-y)^2 D'' +y(1-y)(1-2y)D'  +
\bigg[\delta^2 y^4+\delta\left( \tilde {\beta}-2\delta\right)y^3
&-& \frac{\tilde{\beta}}{4}\left(2+\tilde{ \beta}+12 \delta
\right)
y^2+\nonumber
\\
\frac{1}{4}\left(5( \tilde {\beta}+\delta)^2 +2 \tilde {\beta}(
1+\delta) -4\right)y-\frac{1}{4}(2 \tilde {\beta}+\delta)^2
\bigg]D&=&0.
\label{grosa}
\end{eqnarray}
This is the equation of a spheroidal wave function which has
analytic solution for some particular cases only \cite{murphy}.

We are interested in the asymptotic behaviour of the solutions in
order to define flux near the horizon and far from it. So, let us
study the asymptotic form of (\ref{grosa}).

We first analyse the $y\rightarrow 0$ case. Keeping terms up to
first order, equation (\ref{grosa}) can be written as
\begin{equation}
y^2 D_0 ''+(1-y)yD_0'+\left(-\frac{1}{4}(2\tilde{\beta}+\delta)^
2+
\frac{1}{4}\left(\tilde{\beta}(2+4\delta)-3(\tilde{\beta}^2-
\delta^2-4) \right)y\right)D_0=0,
\end{equation}
whose solution is
\begin{equation}
D_0(y)=P_0\,\,\,y^{-\frac{b}{2}}F\left[ \frac{1}{2}\left(a+b
\right),1-
b,y\right] +Q_0\,\,\,y^{\frac{b}{2}}F\left[
\frac{1}{2}\left(a-b\right),1
+b,y\right],
\label{sol0}
\end{equation}
where $P_0$  and $Q_0$ are  complex constants,
$a=2+\frac{3}{2}(\tilde{\beta}^2-\delta^2)-\tilde{\beta}
(1+2\delta)$ and $b=2\tilde{\beta}+\delta$ and $F[c,d;x]$ is the
confluent hypergeometric function.

Again, the states that define the incoming flux will be a
superposition of this solutions.

In order to discuss the region $y\rightarrow \infty$, we make a
series expansion of (\ref{grosa}) about infinity and discard all
terms of order ${\cal O}(y^{-1})$ and beyond. The resulting
equation is
\begin{equation}
y^2 D_\infty ''+2yD_\infty '+\left(\delta^2\,y^2+\delta\tilde{\beta} \,y-\frac{1}{4} \left(2\tilde{
\beta}+\delta^2+(\tilde{
\beta}+
\delta)(\tilde{ \beta}+3\delta)\right)\right)D_\infty=0,
\end{equation}
whose solutions turn out to be
\begin{eqnarray}
D_{\infty}(y)&=&e^{-i\delta\,y}y^{\frac{1}{2}(a-1)}\bigg(P_\infty \,F\left[\frac{1}{2}\left(1+i
\tilde{\beta}+a\right),1+
a,2i\delta\,y\right]+\nonumber
\\
&&Q_\infty\,U\left[\frac{1}{2}\left(1+i \tilde{\beta}+a\right),1+ a,2i\delta\,y\right]\bigg),
\end{eqnarray}
where
$a=\sqrt{(\tilde{\beta}+1)^2+4\delta(\tilde{\beta}+\delta)}$.
$F[c,d;y]$ and $U[c,d;y]$ are the  confluent hypergeometric
functions discussed previously for the case $\delta =0$.

As we said in the previous section, the regular part of the
solution is given by the $U[c,d;y]$ function. By the same
arguments given there, we can write
\begin{equation}
D_\infty(y)\sim e^{-i\delta y}\sum_{n=0}^{N}\gamma_n\, y^{-n-1},
\end{equation}
where $\gamma_n$ are complex constants. Therefore, the solution
$G(x)$ turns out to be
\begin{equation}
G(x)\sim e^{-i x}\sum _{n=0}^{N}\tilde\gamma_n\, x^{-n-1/2}.
\end{equation}

Finally, let us point out that the previous results can be
understood from a physical point of view as follows. Because of
our choice of radial coordinate, the near horizon region
corresponds to the irregular singular point $x \rightarrow\infty
$. But the horizon is not a physical singularity, and therefore
one should find well behaved solutions near this point. Hence, the
solution can be written as
\begin{equation}
G_{\infty}(x)\sim \frac{1}{x^s}\left(\tilde\alpha_0 +\frac{
\tilde\alpha_1}{x}+\frac{\tilde\alpha_2}{x ^2}+\cdots\right),
\label{general}
\end{equation}
where $s$ is a positive real number. This is just the result found
for $\delta = 0$ and $\delta \neq 0$ with $s=1/2$. Note that the
exponential $e^{ix}$ cannot be obtained with this argument.

To close this section, let us comment on a question that may be
confusing, that is, the fact that the Dirac equation can be {\it
exactly} solved in the {\it regular} rotating BTZ background (see
e.g. \cite{exact}). The difference with the case studied here is
that the redefinition \cite{DAS}
\begin{equation}
r^{2}=r_{+}^{2} cosh^{2}{\mu} - r_{-}^{2}sinh^{2}\mu,
\end{equation}
made to find the exact solution, has no meaning in the limit
$r_{-} \rightarrow r_{+}$, that is, in the extreme BTZ background.
This may be another expression that regular and extreme BTZ black
holes correspond to different physical objects and should be
studied separately.

\section{Ingoing and incoming fluxes and the absorption rate}

As was mentioned before, it is not clear how to prepare an initial
scattering state (incoming flux) in an AdS background because the
solution of the equations of motion at infinity ($r \to \infty$)
are not plane waves. If one needs to find a consistent expression
for the incoming flux at infinity, one could define the incoming
flux by following the procedure developed in \cite{birming} for a
scalar field but this is of no help here, as we will show below.
The problem can be circumvented arguing that the incoming flux,
must be non zero because it represents particles that are sent to
the black hole. Therefore, we only must care to be far enough from
the horizon. We will use here the arguments given in \cite{DAS}
for the case of fermions scattering off a regular BH: since in an
asymptotically flat space $\sqrt{g_{oo}}\rightarrow 1$ at
infinity, and since $\sqrt{g_{oo}}\rightarrow r$ for a BTZ
background, one chooses a point in space a location such that
$\sqrt{g_{oo}}\rightarrow 1$; this occurs for $r\sim l$.

\subsection{The current}

In order to calculate the absorption cross section, it is
necessary to prepare the incoming flux (initial state) and to
obtain the ingoing flux (near $r_{ex}$) later. The current along
$\rho $- direction is
\begin{eqnarray}
j^{\rho } &=&\kappa\bar{\psi}\gamma ^{a}E_{a}^{\rho }\psi,
\\
&=&-\frac{2\kappa}{\ell }\Re \mbox{e}({\cal U}^{\ast }{\cal V}),
\end{eqnarray}
where $\kappa$ is a coupling constant.

From (\ref{ansatz})  and (\ref{ED1}) we obtain
\begin{equation}
j^{\rho }(x) ={\cal A}\left(\frac{x}{
\delta-x}\right)\Re \mbox{e}\left\{i\,G(x)\frac{d}{dx}G^{\ast }(
x)
\right\},
\label{alfincarajo}
\end{equation}
where $\cal A$ is a constant.

\subsection{Flux and cross section}

The black hole absorption rate $\sigma_{abs}$ could be defined as
in quantum mechanics, where it is related to the ratio of the
ingoing flux at the horizon (total number of particles entering
the horizon) and incoming flux at infinity.
\begin{equation}
\sigma _{abs}=\frac{\mathcal{F}\left( x\rightarrow \infty \right)
}{\mathcal{ F}\left( x\rightarrow 0\right) } \label{sigma},
\end{equation}
where
\begin{equation}
\mathcal{F}=\sqrt{-g }\,j^{\rho }(x),
\end{equation}
with  $g$ the determinant of the metric.

It is straightforward to prove that
\begin{equation}
\mathcal{F}={\cal A}\,\sqrt{r_{ex}^2+\frac{\alpha\ell
r_{ex}}{x}}\left( \frac{x}{1-x}\right) \Re \mbox{e}\left\{i\,G(x)
\frac{d}{dx}G^{*}\left(x\right)\right\} \label{otra}.
\end{equation}

Now we must evaluate this quantity for $\delta =0$ and $\delta
\neq 0$, near the horizon and at spatial infinity. As we showed in
the previous section, however, there is  a general form of the
solution given by (\ref{general}), as we showed in the previous
section for the near horizon region.

It is straightforward to evaluate the flux near the horizon:
\begin{eqnarray}
\mathcal{F}&=&{\cal A}\,\sqrt{r_{ex}^2+\frac{\alpha\ell \,r_{ex}}{x}} \,\frac{x}{1-x}\, \Re
\mbox{e}\left\{\frac{i}{x
^{2s}}\left( \tilde{\alpha}_0 +\frac{ \tilde{\alpha}_1}{x}+\frac{\tilde{\alpha}_2}{x^2}+ \cdots
\right) \left(\kappa _0 +
\frac{\kappa _1}{x}+\frac{\kappa _2}{x^2}\cdots\right) \right\}\nonumber
\\
&=&{\cal A}\,\sqrt{r_{ex}^2+\frac{\ell^2}{x}}
\,\frac{1}{1-x}\, \Re \mbox{e}\left\{\frac{i}{x^{2s-1}}\left(\tilde{\alpha_0}\kappa_0+\frac{
\tilde{\alpha}_0\kappa_1+
\tilde{\alpha}_1\kappa_0}{x}+\frac{\tilde{\alpha}_0\kappa_2+\tilde{\alpha}_1\kappa_1+\tilde{
\alpha}_2\kappa_0}{x^
2}\cdots\right)\right\}.\nonumber
\end{eqnarray}
Here $\tilde{\alpha}_i$  and
$\kappa_0=i\tilde{\alpha}_0^{*}, \kappa_1=i\tilde{\alpha}_1^{*}-s \tilde{\alpha}_0^{*},
\kappa_2=
i\tilde{\alpha}_2^{*}-\tilde{\alpha} _1^{*} - s \tilde{\alpha}_1^{*},\cdots $, are numerical
constants.

The last expression shows that for $s\geq 1/2$
\begin{equation}
\lim_{x\rightarrow\infty}{\cal F}\to 0.
\end{equation}

In order to define the incoming flux in a consistent way, one
could define the ingoing and outgoing states as a complex linear
combination of the asymptotic solutions at infinity, as done in
\cite{birming} for a scalar field. In the present case, ($\delta
=0$) this means to consider

\begin{equation}
G_{r\rightarrow \infty }(x)=e^{-ix}x^{\frac{\widetilde{\beta }+1}{2}}(PF[1+%
\frac{1}{2}(1+i)\widetilde{\beta },2+\widetilde{\beta };2ix]\pm QU[1+\frac{1%
}{2}(1+i)\widetilde{\beta },2+\widetilde{\beta };2ix]),
\label{infinitylfux}
\end{equation}
where $P$ y $Q$ are complex constants and the positive (negative)
sign corresponds to the ingoing (outgoing) waves. It is
straightforward to show that, by replacing this expression into
Eq. (\ref{otra}), the flux diverges (to leading order) as
$x^{-\frac{1}{2}}$ at infinity, a result that is valid for $\delta
= 0$  as well as $\delta \neq 0$.

Given that the previously defined flux is divergent, we can use
the argument found in \cite{DAS} in order to define the incoming
flux. Choosing $x\sim \it{const.}\, l$ (corresponding to $r\sim
l$) we find
\begin{equation}
\lim_{r\rightarrow l}{\cal F}\to \mbox{constant}.
\end{equation}

Finally, from (\ref{sigma}) one proves that
\begin{equation}
\sigma_{abs}=0.
\end{equation}
for the extreme 2+1 dimensional black hole.

Let us point out that this result has no contradictions with the
non-extreme black hole. The reason is that our argument about the
form of the function $G(x)$ is still valid in the regular black
hole background, but the damping factor in front of
(\ref{alfincarajo}) is different for such case because Eqs.
(\ref{ED1}) and (\ref{ED2}) are not the same for the extreme and
non-extreme cases and hence one cannot map one into the other in a
continuous manner.

\subsection{Reciprocal space}

In \cite{gm} we argued that it is possible to circumvent the
problem of defining the flux of particles at infinity in a AdS
spacetime by working in a sort of reciprocal space, which means to
map the original problem into another one, where the concept of
free particles at infinity holds.

In the present case this  can also be performed. The arguments
given before --in \cite{gm}-- are still valid and much of the
discussion in this section is based on it.

The main issue is to write the equation (\ref{eqg}) in its
canonical form
\begin{equation}
u''(x) + I(x)u(x)=0,
\end{equation}
where $I(x)$ is the invariant.

Again, Ii our problem we must distinguish two cases: $\delta=0$
and $\delta\neq 0$.

For $\delta=0$, one defines
\begin{equation}
G(x)=\frac{u_0 (x)}{\sqrt{x}},
\end{equation}
where $u_0 (x)$ satisfies the equation
\begin{equation}
u_0''+\left(1+\frac{\tilde{\beta}}{x}-\frac{\tilde{\beta}(\tilde{\beta}+2)}{4\,x^2}\right)u_0=0.
\label{u0}
\end{equation}

For $\delta \neq 0$, defining $\xi=x/\delta$ and
\begin{equation}
G(\xi)=\frac{\sqrt{\xi-1}}{\xi}\,u_1 (\xi),
\end{equation}
one finds that $u_1 (\xi)$ satisfies
\begin{equation}
u_1''+\left(\delta^2+\frac{\tilde{\beta}(2\delta+1)-1}{2\,\xi}+\frac{(\tilde{\beta}+2\delta)^2-1}{
4\,\xi^2}+\frac{1-\tilde{
\beta}}{1-\xi}+\frac{(3/4}{(\xi-1)^2}\right)u_1=0.
\label{u1}
\end{equation}

In (\ref{u0}) and (\ref{u1}), $x$ and $\xi$ play the role of
radial coordinates, respectively, in a Schr\"odinger-type
equation. The reciprocal spaces are then ${\cal H}_0=\{\phi,x\}$
and ${\cal H}_1=\{\phi,\xi\}$, where $\phi$ is an angular
coordinate.

Additionally, one must impose that
\[
u_0(x=0)=0=u_1(\xi=0),
\]
in order to have continuous solutions everywhere
\cite{elipticos}.

Finally, one note that the {\it potential} terms in (\ref{u0}) and
(\ref{u1}) vanish as $x\rightarrow\infty$ and $\xi
\rightarrow\infty$, respectively. Therefore, it is possible to
define asymptotic states as in the usual scattering theory;
meaning solutions like $A_0(\phi)e^{ix}$ for $u_0$ and
$A_1e^{i\delta\xi}$ for $u_1$, where $A_i$  are the scattering
amplitudes and the exponential terms, are the asymptotic states.

However, invoking the optical theorem, one can directly prove that
this gives rise to a vanishing total cross section (see \cite{gm}
for details). This implies a zero absorption cross section as was
calculated in the previous section when returning to the original
space

\section{Conclusions}
In this work we have discussed the problem of fermions scattering
off a 2+1-dimensional extreme black hole background.

We showed that the Dirac equation can be solved when the {\it fine
tuning condition} is satisfied.

We also gave a general argument in order to show that the
solutions of the Dirac equation near the horizon need to have a
precise form, what was checked by solving them.

With these solutions we were able to prove that the flux of
incoming particles near the horizon vanishes and therefore that
the absorption cross section becomes zero, which agrees with
previous results for scalar particles.

The previous result is not contradictory with the well known fact
that for a regular BH background, the cross section is
proportional to the area (entropy) of the horizon. In fact, our
result is consistent with the fact that both solutions of the
Einstein equations (the regular and the extreme black holes)
define different topologies of the spacetime.

We also proved that the description in terms of the {\it
reciprocal space} yields the same results implying that it could
be considered as the space where the scattering in AdS should be
defined.

Finally, let us remark that our main  result agrees with arguments
that show that the extreme black hole entropy is zero and
therefore can be considered as fundamental objects.

 \acknowledgments

This work was supported by UCV-VRIEA-DI under Grant N$^{0}$
123.763/2002 (SL)and FONDECYT under Grants N$^{0}$ 3000005 (FM)
and N$^{0}$ 1020061(LV). One of us (SL) acknowledges the
Departamento de F\'{\i}sica, Universidad de Santiago de Chile for
hospitality. LV and FM   would like to thanks M. Ba\~nados for
enlightening discussions. JS and   FM would like to thank  J.
Zanelli and R. Troncoso for several discussions. FM acknowledges
also useful discussions with J. Gamboa who suggested this problem.
We thank Dr. U. Raff for a careful reading of the manuscript.

\end{document}